\documentclass[twocolumn,preprintnumbers,amsmath,amssymb]{revtex4}
\usepackage{graphicx}
\usepackage{dcolumn}
\usepackage{textcomp}

\begin{document}

\preprint{}

\title{Method for Observing Intravascular BongHan Duct}

\author{Xiaowen Jiang}
 \altaffiliation{A permanent address: Dept. of Veterinary, College of Agriculture, Yanbian University, Jilin, China}
\author{Byung-Cheon Lee}
\author{Chunho Choi}
\author{Ku-Youn Baik}
\author{Kwang-Sup Soh}
 \altaffiliation{Correspondence: Kwang-Sup Soh}
 \email{kssoh@kmc.snu.ac.kr}
 \homepage{http://kmc.snu.ac.kr/~kssoh}
\affiliation{Biomedical Physics Lab., School of Physics, Seoul National University, Seoul, 151-747, Korea}

\author{Hee-Kyeong Kim}
\affiliation{School of Chemical Engineering, Seoul National University}

\author{Hak-Soo Shin}
\affiliation{Department of Physics Education, Seoul National University}

\author{Kyung-Soon Soh}
\affiliation{College of Oriental Medicine, Sae Myong University, Chungbook Korea}

\author{Byeung-Soo Cheun}
\affiliation{School of Biochemistry, Inha University, Inchon, Korea}

\date{\today}

\begin{abstract}
A method for observing intra blood vessel ducts which are threadlike bundle of 
tubules which form a part of the BongHan duct system. By injecting 10\% dextrose
solution at a vena femoralis one makes the intravascular BongHan duct thicker 
and stronger to be easily detectable after incision of vessels. 
The duct is semi-transparent, soft and elastic, and composed of smaller tubules 
whose diameters are of 10$\mu$m order, which is in agreement with BongHan theory.
\end{abstract}

\maketitle

\section{ntroduction}
Recently observation of hidden anatomical structures of threadlike ducts inside blood vessels was reported.\cite{Soh} These threads of tubular bundles are intravascular BongHan ducts.

According to the BongHan theory \cite{Kim,KimBH} there is a circulatory system that is completely different from blood systems, lymphatic systems, and nervous systems. The system is composed of BongHan ducts of which acupuncture meridian system of Traditional Korean Medicine is a subsystem. In addition to the meridians there is a host of network of BongHan ducts in which BongHan liquid flows, and the liquid is quite distinct from blood, or other known fluids inside animals or humans.

The BongHan theory claimed that some parts of BongHan ducts run inside arteries and veins as threadlike bundles of tubules floating in the flow of blood. However, no one has confirmed it despite intensive search for it in 1960s, and thus the theory itself has been disregarded for almost forty years. The main reason for not reproducing BongHan's results is that the staining material and method, the key technique in BongHan theory, were not disclosed.

In this paper we present a completely new method which is distinctively different from BongHan's staining approach. Our method is a relatively simple procedure to disclose intra blood vessel ducts (IBVD) which form part of the net of BongHan ducts. We have not yet developed to search other ducts such as inside lymph vessels and  nerves, and under the skin. Nevertheless our new finding is easily reproducible by others, and will usher to further studies of the BongHan duct system such as histological, chemical, and physiological analysis.

\section{Methods}
Our principal contribution is the discovery of the new method to reveal and take samples of BongHan ducts inside arteries and veins. These IBVDs were found in various large blood vessels of rats and rabbits. We examined successfully ten rats (Sprague Dawley, бн 900g) and four rabbits (New Zealand White, бн2000g) which were obtained from the Laboratory Animal Center of Seoul National University with ACUC permission for studying acupuncture meridian systems. The following is a representative example of surgery operation with a rat, and the time varies depending upon the subject animal and the details of operation. The procedure for rabbits is similar.

An intravenous injection of 10\% dextrose solution at the left vena femoralis is started after anesthetization. At 30th minute about 35ml dextrose is injected, and the jugular vein is cut to bleed by itself. At 55th minute the frontal side of the rat is incised, and internal organs, flesh, and side muscles are removed so that blood systems are exposed for operational ease. Around this time the dextrose injection is stopped. At 105th minute after beginning of the dextrose injection the search of the IBVD starts from the inferior vena cava and lasts more than two hours taking samples from various blood vessels.

We do not understand the mechanism of the 10\% dextrose solution which seems to help the IBVD to become thicker, and stronger so that it is easily noticeable and to be stretched out long. Without the dextrose solution the IBVD is too thin and weak to be seen.

The IBVD is buried inside a string of fibrins with blood coagulated around it, and becomes thick, which is the reason we can find it inside the opened vessel. The blood clot is dissolved by urokinase treatment. One vial of fibrinolytic agent urokinase 100,000 I.U. is dissolved in 100ml saline water. The IBVD with blood clot is immersed into the urokinase solution at about 37\textcelsius\ for two or three hours to obtain the IBVD inside the fibrin string.

\section{Results}
  Basic properties of the IBVD are in agreement with BongHan theory.\cite{Kim,KimBH} They are soft, elastic, and semi-transparent. Their diameters are about 50$\mu$m. The IBVD has many smaller tubules like a bundle bound with fiber. The small tubule is about 10$\mu$m in diameter, and is stained well with methylene blue (0.02\%) 
\begin{figure}
\resizebox{8cm}{!}{
\includegraphics{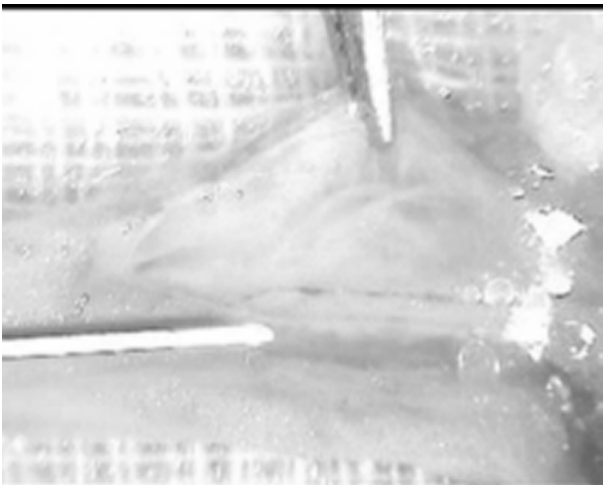}}
\caption{\label{Duct}An intravascular duct in an inferior vena cava}
\end{figure}

\begin{figure}
\resizebox{8cm}{!}{
\includegraphics{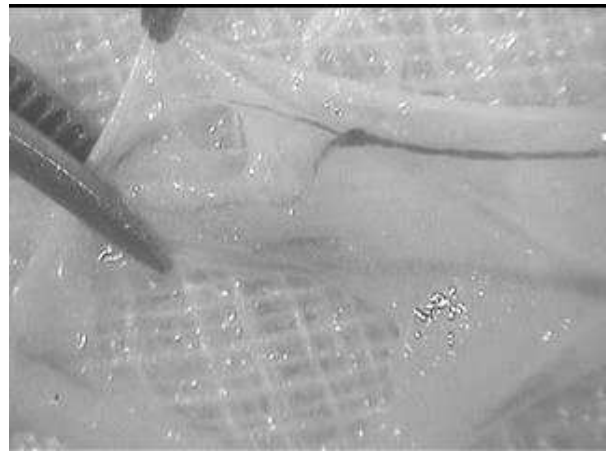}}
\caption{\label{Bifurcation}Bifurcation of a BongHan duct}
\end{figure}

An example of IBVD coagulated with fibrin inside the incised inferior vena cava is shown in Fig.\ref{Duct} The IBVD starts from above the tip of the syringe needle. (gauge 23, 5ml/cc), and is parallel with the needle

Fig.\ref{Bifurcation} shows that an IBVD bifurcates to two IBVDs where the inferior vena cava branches to the left and right iliac veins. This branching is in complete agreement with BongHan theory. 

\begin{figure}
\resizebox{8cm}{!}{
\includegraphics{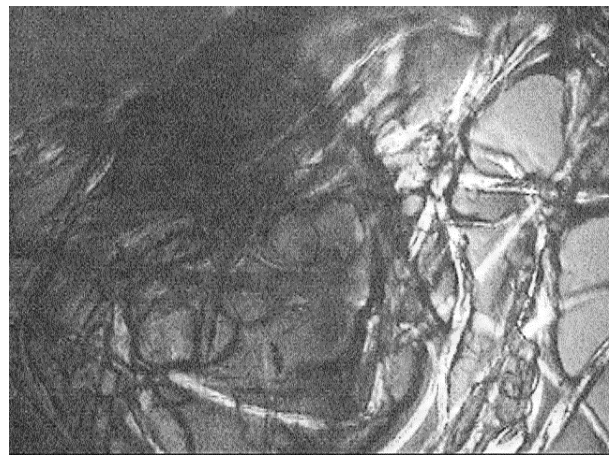}}
\caption{\label{Entangle}Entangled BongHan ducts inside atrium}
\end{figure}

 Entangled BongHan tubules which was found inside atrium is shown in Fig.\ref{Entangle}. BongHan theory also claims that there are multiple of IBVDs inside a heart.

We observed an IBVD inside the abdominal aorta that is hold by several two-threads stretched out to the vessel wall like a ladder. The distances between the rungs are about 1.5cm. Those side threads might be smaller IBVDs or simply attaching fibers which requires further studies.

\section{Discussion}
 It is an immediate question that why the IBVD is not found in surgeries which have been done millions of times. The main reason is that the IBVD itself is too thin, semi-transparent, and weak to be detected or captured unless special care is taken. In an ordinary surgery a divided IBVD shrivels and is enshrouded with blood clots which can not be restored unless treated by urokinase or similar material.

Our method is to inject 10\% dextrose solution which makes the IBVD thicker and stronger by fibrins of coagulated blood around it. The IBVD may be a 'seed' for fibrins to gather around it, and dextrose may enhance this effect. It requires further studies to find the mechanism of this phenomenon. Another effect of the dextrose solution is to dilute blood inside the vessel so that the blood clotted IBVD is more noticeable because the diluted blood becomes almost transparent. Without such diluted blood the IBVD can not be seen easily.

It is possible to mistake a string of coagulated fibrins for the IBVD when taking samples. The fibrin-string does not have tubular substructures. Furthermore it is very improbable that the fibrin-string without the 'seed' of the IBVD branches to two pieces at the same anatomical positions for all the rats.

The limitation of our method is that it applies to the BongHan ducts only inside blood vessels. Nevertheless, this confirmation of BongHan theory is the first one ever achieved, and will open vast areas of work for biological and medical research, and Traditional Korean Medicine. Taking just one example, we might wonder what is the effect of surgery when BongHan ducts are left cut and disconnected.

\begin{acknowledgments}
Soh(SNU) thanks Professor Marco Bischof for the reference 3, and is indebted to M.D. Do-Hyun Kim, V.M.D.s Dae-Hun Park and Chung-Hyun Lee, and O.M.D. Jong-Han Lee. 

This work is supported in part by BK21 KRF, KOSEF, and Ministry of Industry and Resources, NongHyup SNU branch, and Samsung Advanced Institute of Technology.
\end{acknowledgments}

\bibliographystyle{abbrv}
\bibliography{bonghan}

\begin{thebibliography}{1}

\bibitem{Soh}
X.~Jiang, H.-K. Kim, H.-S. Shin, B.-C. Lee, C.~Choi, K.-S. Soh, B.-S. Cheun,
  K.~yoon Baik, and K.-S. Soh.
\newblock Threadlike bundle of tubules running inside blood vessels: New
  anatomical structure.
\newblock August 2002.

\bibitem{KimBH}
B.~Kim.
\newblock {\em On the Kyungrak System}.
\newblock Medical Science Press, Pongyang, Korea, 1963.

\bibitem{Kim}
B.~Kim.
\newblock Theory of kyungrak.
\newblock {\em J. The Academy of Medical Sciences of the Democratic People's
  Republic of Korea}, pages 9--54, 1965.

\end{thebibliography}
\end{document}